\documentclass[12pt]{article}
\usepackage{oldgerm}
\usepackage{yfonts}
\usepackage{graphics}
\usepackage{bm}
\usepackage{graphicx}
\usepackage{amsmath}
\usepackage{amssymb}
\usepackage{amstext}
\usepackage{amscd}
\usepackage{amsfonts}
\usepackage{appendix}
\usepackage{vmargin}
\usepackage{epstopdf}
\usepackage{color}
\usepackage{hyperref}
\DeclareMathAlphabet{\EuFrak}{U}{euf}{m}{n}
\DeclareMathAlphabet{\EuScript}{U}{eus}{m}{n}

\newcommand{\nd}{\noindent}

\newcommand{\be}{\begin{equation}}
\newcommand{\ee}{\end{equation}}
\newcommand{\ben}{\begin{eqnarray}}
\newcommand{\een}{\end{eqnarray}}

\title{{\bf Gupta-Feynman based Quantum Field Theory of Einstein's Gravity}}
\author{{{\bf A. Plastino $^{\boldsymbol{1,3,4}}$, M.C.Rocca$^{\boldsymbol{1,2,3}}$}}, \\
\small{$^1$ Departamento de F\'{\i}sica,
Universidad Nacional de La Plata,}\\
\small{$^2$ Departamento de Matem\'{a}tica,
Universidad Nacional de La Plata,}\\
\small{$^3$ Consejo Nacional de Investigaciones Cient\'{\i}ficas
y Tecnol\'{o}gicas}\\
\small{(IFLP-CCT-CONICET)-C. C. 727, 1900 La Plata -
Argentina}\\\small{$^4$  SThAR - EPFL, Lausanne, Switzerland}}

\setpapersize{A4}
\setmargins{2cm}       
{1.5cm}                        
{17cm}                      
{23.42cm}                    
{10pt}                           
{1cm}                           
{0pt}                             
{2cm}                           

\date{\today}

\begin{document}

\maketitle

\vspace{-5mm}

\begin{abstract}

\nd This paper is an {\sf application} to Einstein's gravity (EG) of the
mathematics developed in A. Plastino, M. C. Rocca:
J. Phys. Commun. {\bf 2}, 115029 (2018). We will quantize EG by appeal to the most general quantization approach, 
the Schwinger-Feynman variational principle,  which is more appropriate and rigorous that the functional integral method,
when we are in the presence of derivative couplings
\nd  We base our efforts  on works by Suraj N. Gupta and Richard P. Feynman so as to
undertake the construction of a Quantum Field Theory (QFT) of
Einstein Gravity (EG).  We explicitly use the Einstein Lagrangian elaborated
by Gupta \cite{g1} but choose a new constraint for the theory
that differs from Gupta's one.  In this way, we avoid the 
problem of lack   of  unitarity for the  $S$ matrix that afflicts the procedures of
Gupta and Feynman.\\
Simultaneously, we significantly simplify the handling of constraints. This
 eliminates the need to appeal to ghosts for guarantying  the unitarity
of the theory.\\
Our ensuing approach  is obviously non-renormalizable. However, this 
inconvenience can be overcome by appealing tho the mathematical theory
developed by Bollini et al. \cite{tp3,tp18,tp19,tp20,pr}\\
Such developments were founded in the works of Alexander Grothendieck \cite{gro}
and in the theory of Ultradistributions of Jose Sebastiao e Silva \cite{tp6}
(also known as Ultrahyperfunctions).\\
Based on these works,  we have constructed a mathematical
edifice, in a lapse of about 25 years, that is able to quantize non-renormalizable Field Theories (FT).\\
Here we specialize this mathematical theory to treat the quantum field theory of Einsteins's gravity (EG).\\ 
Because we are using a Gupta-Feynman inspired EG Lagrangian, we are
able to evade the intricacies of Yang-Mills theories.\\ 
\nd
\nd

\end{abstract}

\newpage

\tableofcontents

\newpage

\renewcommand{\theequation}{\arabic{section}.\arabic{equation}}

\section{Introduction}

\nd Quantifying Einstein gravity (EG) is still an open problem, a kind of holy grail for quantum field theory (QFT).
 The failure of some attempts in this direction have failed because i) they appeal   to  Rigged Hilber Space (RHS) 
 with undefined metric, ii) problems of non-unitarity, and also iii) non-renormalizablity issues. Here we quantize EG by appeal to the most general quantization approach, 
the Schwinger-Feynman variational principle, which is more appropriate and rigorous that
the functional integral method, when we are in the presence of derivative couplings.\\
\nd Here we build up an unitary  EG's  QFT in the wake 
of related  effort by Suraj N. Gupta \cite{g1}. We deviate from his work by using a different EG-constraint, facing then 
a  problem similar to that posed by Quantum Electrodynamics (QED). In order to quantize the  
concomitant non-renormalizable  variational  
 problem we appeal to mathematics developed by Bollini et al.
\cite{tp3,tp18,tp19,tp20,pr}, based upon the theory 
of  Ultradistributions de J. Sebastiao e Silva (JSS) \cite{tp6},
also known as  Ultrahyperfunctions. 
 The above cited mathematics were  specifically devised to 
quantify non-renormalizable field  theories during 25 years, culminating in \cite{pr}. 
We consequently face a theory similar to QED, endowed with unitarity at all finite 
orders in the power expansion  in $G$  (gravitation constant) of the EG Lagrangian. 
This was attempted without success  first by  Gupta and then by  Feynman,  in his celebrated  
Acta Physica Polonica paper \cite{fe}.\\

\nd Mathematically, quantizing a non-renormalizable field theory is tantamount to suitably  
defining the product of two distributions (a  product in a ring with 
zero-divisors in configuration space), an old problem in functional theory tackled 
successfully in  \cite{tp3,tp18,tp19,tp20,pr}.\\
Remark that, in QFT, the problem of evaluating the product
of distributions with coincident point singularities is related 
to the asymptotic behavior of loop integrals
of propagators.\\ 

\nd In  references \cite{tp3, tp18, tp19, tp20} it was demonstrated
that it is possible to define a general convolution between the
ultradistributions of JSS \cite{tp6} (Ultrahyperfunctions).
This convolution yields another Ultrahyperfunction.
Therefore, we have a product in a ring with zero divisors.
Such a ring is the space of
distributions of exponential type, or ultradistributions of
exponential type, obtained applying the anti-Fourier transform to
the space of tempered ultradistributions or ultradistributions of
exponential type.\\

\nd  We must clarify at this point that the ultrahyperfunctions
are the generalization and extension to the complex plane
of the Schwartz tempered distributions and the
distributions of exponential type. That is, the
tempered distributions and those of exponential type
are a subset of the ultrahyprefunctions.\\

\nd In our work we do not use counter-terms to get rid of infinities, 
because our convolutions are always finite.  
{\it We do not want counter-terms, since a non-renormalizable 
theory involves an infinite 
number of them }. \\

\nd At the same time, we conserve 
all extant solutions to the problem of running coupling constants
and the renormalization group. The convolution,  once obtained, converts configuration space 
into a ring with zero-divisors. In it,   one has now defined a product between the ring-elements. 
Thus, any unitary-causal-Lorentz invariant  theory quantized in such a manner
 becomes  predictive. The distinction  between renormalizable on non-renormalizable QFT's 
 becomes unnecessary now. \\

\nd With our convolution, that uses Laurent's expansions in the parameter employed to define it,
all finite constants of the  convolutions become completely determined, eliminating arbitrary choices 
of finite constants. This is tantamount to eliminating  all 
finite renormalizations of the theory. The independent term in the  
Laurent expansion give the convolution value.  
This translates to configuration  space
the product-operation  in a ring
with divisors of zero.\\

\nd This paper is organized as follows:
\begin{itemize}
\item Section 2 presents preliminary materials.\\
\item Section 3 is devoted to the QFT Lagrangian for EG\\
\item In Section 4 we quantize the ensuing theory.\\
\item In Section 5 the graviton's self-energy is evaluated up to
second order.\\
\item In Section 6 we introduce axions into our picture and deal with the axions-gravitons interaction.\\
\item In Section 7 we calculate the graviton's self-energy in the
presence of axions.\\
\item In Section 8 we evaluate, up to second order, the axion's
self-energy.\\
\item Finally, in Section 9, some conclusions are drawn.\\
\end{itemize}

\section{Preliminary Materials}

\setcounter{equation}{0}

\nd We appeal here  to the most general quantification approach, 
Schwinger-Feynman variational principle \cite{vis},
which is able  to deal even with high order supersymmetric theories , as exemplified by 
\cite{prasad,wz}. Such theories can not be quantized with the usual Dirac-brackets technique. \\

\nd  We introduce the action for a set of fields defined by
\begin{equation}
\label{ep2.1}
\boldsymbol{\cal S}[\sigma(x),\sigma_0,\phi_A(x)]=
\int\limits_{\sigma_0}^{\sigma(x)}{\cal L}[\phi_A(\xi),\partial_\mu\phi_A(\xi),\xi]d\xi,
\end{equation}
where $\sigma(x)$ if a space-like surface   passing through the point 
 $x$.  $\sigma_0$  is that surface at the remote past, at which all field variations vanish. The  
 Schwinger-Feynman variational principle dictates that \\

\nd ''Any Hermitian infinitesimal variation $\delta\boldsymbol{\cal S}$ of the action
induces a canonical transformation of the vector space in which the quantum system
is defined, and the generator of this transformation is this same operator $\delta\boldsymbol{\cal S}$''.\\

\nd Accordingly, the following equality holds:\\
\begin{equation}
\label{ep2.2}
\delta\phi_A=i[\delta\boldsymbol{\cal S},\phi_A].
\end{equation}
\nd Thus, for a  Poincare transformation we have
\begin{equation}
\label{ep2.3}
\delta\boldsymbol{\cal S}=a^{\mu}\boldsymbol{\cal P}_{\mu}+
\frac {1} {2}a^{\mu v}\boldsymbol{\cal M}_{\mu v},
\end{equation}
\nd where the field variation is given by 
\begin{equation}
\label{ep2.4}
\delta\phi_a=a^{\mu}\hat{P}_{\mu}\phi_A+\frac {1} {2}a^{\mu v}\hat{M}_{\mu v}\phi_A.
\end{equation}
\nd From  (\ref{ep2.2}) one gathers that 
\begin{equation}
\label{ep2.5}
\partial_\mu\phi_A=i[\boldsymbol{\cal P}_\mu,\phi_A].
\end{equation}
\nd Specifically, 
\begin{equation}
\label{ep2.6}
\partial_0\phi_A=i[\boldsymbol{\cal P}_0,\phi_A].
\end{equation}
\nd This last result will be employed in quantizing  EG.

\section{The Lagrangian of Einstein's QFT}

\setcounter{equation}{0}

\nd Our EG Lagrangian reads \cite{g1}
\begin{equation}
\label{ep3.1}
{\cal L}_G=\frac {1} {\kappa^2}\boldsymbol{R}\sqrt{|g|}-\frac {1} {2}
\eta_{\mu v}\partial_\alpha h^{\mu\alpha}
\partial_\beta h^{v\beta},
\end{equation}
 where 
$\eta^{\mu\nu}=diag(1,1,1,-1)$, $h^{\mu\nu}=\sqrt{|g|}g^{\mu\nu}$
\nd The second term in eq. (\ref{ep3.1}) fixes the gauge.
 We apply now the linear approximation
\begin{equation}
\label{ep3.2}
h^{\mu v}=\eta^{\mu v}+\kappa\phi^{\mu v},
\end{equation}
\nd where  $\kappa^2$ is the gravitation's constant and  $\phi^{\mu v}$
the graviton field. We write
\begin{equation}
\label{ep3.3}
{\cal L}_G={\cal L}_L+{\cal L}_I,
\end{equation}
\nd where
\begin{equation}
\label{ep3.4}
{\cal L}_L=-\frac {1} {4}[\partial_\lambda\phi_{\mu v}\partial^\lambda\phi^{\mu v}-2
\partial_\alpha\phi_{\mu\beta}\partial^{\beta}\phi^{\mu\alpha}+2
\partial^{\alpha}\phi_{\mu\alpha}\partial_\beta\phi^{\mu\beta}],
\end{equation}
\nd and, up to 2nd order,  one has \cite{g1}: 
\begin{equation}
\label{ep3.5}
{\cal L}_I=-\frac {1} {2}\kappa\phi^{\mu v}[\frac {1} {2}\partial_\mu\phi^{\lambda\rho}
\partial_v\phi_{\lambda\rho}+\partial_\lambda\phi_{\mu\beta}\partial^\beta\phi_v^\lambda-
\partial_\lambda\phi_{\mu\rho}\partial^\lambda\phi_v^\rho],
\end{equation}
having made use of the constraint
\begin{equation}
\label{ep3.6}
\phi_\mu^\mu=0.
\end{equation}
 This constraint is required in order to satisfy  gauge invariance  \cite{kle}
 For the  graviton we have then
\begin{equation}
\label{ep3.7}
\square\phi_{\mu v}=0, 
\end{equation}
\nd whose solution is 
\begin{equation}
\label{ep3.8}
\phi_{\mu v}=\frac {1} {(2\pi)^{\frac {3} {2}}}\int\left[\frac {a_{\mu v}(\vec{k})} {\sqrt{2k_0}}
e^{ik_\mu x^\mu}+\frac {a^+_{\mu v}(\vec{k})} {\sqrt{2k_0}}
e^{-ik_\mu x^\mu}\right]d^3k,
\end{equation}
\nd with $k_0=|\vec{k}|$.

\section{The Quantization of the Theory}

\setcounter{equation}{0}

\nd We need some definitions. The energy-momentum tensor is 
\begin{equation}
\label{ep4.1}
T_\rho^\lambda=\frac {\partial{\cal L}} {\partial\partial^\rho\phi^{\mu v}}\partial^\lambda\phi^{\mu v}-
\delta_\rho^\lambda{\cal L},
\end{equation}
 and the time-component of the four-momentum  is 
\begin{equation}
\label{ep4.2}
\boldsymbol{\cal P}_0=\int T_0^0\;d^3x.
\end{equation}
\nd Using  (\ref{ep3.4}) we have
\[T_0^0=\frac {1} {4}[\partial_0\phi_{\mu v}\partial^0\phi^{\mu v}+
\partial_j\phi_{\mu v}\partial^j\phi^{\mu v}-2
\partial_\alpha\phi_{\mu 0}\partial^0\phi^{\mu\alpha}-2
\partial_\alpha\phi_{\mu j}\partial^j\phi^{\mu\alpha}+\]
\begin{equation}
\label{ep4.3}
2\partial_{\alpha}\phi^{\mu\alpha}\partial_0\phi_\mu^0+
2\partial_{\alpha}\phi^{\mu\alpha}\partial_j\phi_\mu^j].
\end{equation}
 Consequently, 
\begin{equation}
\label{ep4.4}
\boldsymbol{\cal P}_0=\frac {1} {4}\int |\vec{k}|\left[
a_{\mu v}(\vec{k})a^{+\mu v}(\vec{k})+    
a^{+\mu v}(\vec{k})a_{\mu v}(\vec{k})\right]d^3k.
\end{equation}
 Appeal to  (\ref{ep2.6}) leads to 
\[[\boldsymbol{\cal P}_0,a_{\mu v}(\vec{k})]=-k_0a_{\mu v}(\vec{k})\]
\begin{equation}
\label{ep4.5}
[\boldsymbol{\cal P}_0,a^{+\mu v}(\vec{k})]=k_0a^{+\mu v}(\vec{k}). 
\end{equation}
\nd From the last relation in  (\ref{ep4.5}) one gathers that 
\begin{equation}
\label{ep4.6}
|\vec{k}|a^{+\rho\lambda}(\vec{k^{'}})=
\frac {1} {2}\int |\vec{k}|
[a_{\mu v}(\vec{k}),a^{+\rho\lambda}(\vec{k^{'}})]a^{+\mu v}(\vec{k})\;d^3k.
\end{equation}
\nd The solution of this integral equation is 
\begin{equation}
\label{ep4.7}
[a_{\mu v}(\vec{k}),a^{+\rho\lambda}(\vec{k^{'}})]=
\left[\delta_\mu^\rho\delta_v^\lambda+\delta_v^\rho\delta_\mu^\lambda\right]
\delta(\vec{k}-\vec{k^{'}}).
\end{equation}
As customary, the physical state  $|\psi>$  of the theory is defined via the equation
\begin{equation}
\label{ep4.8}
\phi_\mu^\mu|\psi>=0.
\end{equation}
\nd We use now the the usual definition 
\begin{equation}
\label{ep4.9}
\Delta_{\mu\nu}^{\rho\lambda}(x-y)=<0|T[\phi_{\mu\nu}(x)\phi^{\rho\lambda}(y)]|0>.
\end{equation}
\nd The graviton's  propagator then turns out to be 
\begin{equation}
\label{ep4.10}
\Delta_{\mu\nu}^{\rho\lambda}(x-y)=\frac {i} {(2\pi)^4}
(\delta_\mu^\rho\delta_v^\lambda+\delta_v^\rho\delta_\mu^\lambda)
\int\frac {e^{ik_\mu(x^\mu-y^\mu)}} {k^2-i0}d^4k.
\end{equation}
As a consequence, we can write
\begin{equation}
\label{ep4.11}
\boldsymbol{\cal P}_0=\frac {1} {4}\int |\vec{k}|\left[
a_{\mu v}(\vec{k})a^{+\mu v}(\vec{k^{'}})+    
a^{+\mu v}(\vec{k^{'}})a_{\mu v}(\vec{k})\right]
\delta(\vec{k}-\vec{k^{'}})d^3k d^3k^{'},
\end{equation}
or
\begin{equation}
\label{ep4.12}
\boldsymbol{\cal P}_0=\frac {1} {4}\int |\vec{k}|\left[    
2a^{+\mu v}(\vec{k^{'}})a_{\mu v}(\vec{k})+
\delta(\vec{k}-\vec{k^{'}})\right]
\delta(\vec{k}-\vec{k^{'}})d^3k d^3k^{'}.
\end{equation}
Thus, we obtain
\begin{equation}
\label{ep4.13}
\boldsymbol{\cal P}_0=\frac {1} {2}\int |\vec{k}|    
a^{+\mu v}(\vec{k})a_{\mu v}(\vec{k})d^3k,
\end{equation}
where we have used the fact that the product of two deltas with the same argument vanishes \cite{tp3}, i.e., 
$\delta(\vec{k}-\vec{k^{'}}) \delta(\vec{k}-\vec{k^{'}})=0$.
 This illustrates the fact that using  Ultrahyperfunctions 
 is here equivalent to adopting the normal order in the definition of the time-component of the four-momentum
\begin{equation}
\label{ep4.14}
\boldsymbol{\cal P}_0=\frac {1} {4}\int |\vec{k}|:\left[
a_{\mu v}(\vec{k})a^{+\mu v}(\vec{k})+    
a^{+\mu v}(\vec{k})a_{\mu v}(\vec{k})\right]:d^3k.
\end{equation}
\nd Now, we must insist  on the fact that the physical state should satisfy
 not only Eq.  (\ref{ep4.8}) but also the relation  (see \cite{g1})
\begin{equation}
\label{ep4.15}
\partial_\mu\phi^{\mu v}|\psi>=0.
\end{equation}
 The ensuing theory is similar to the QED-one obtained via the quantization approach of Gupta-Bleuler. This implies that 
 the theory is unitary for any finite perturbative order. In this theory only one type of graviton emerges, 
 $\phi^{12}$, while in  Gupta's approach two kinds of graviton emerge. 
 Obviously, this happens for a non-interacting   theory, as remarked by  Gupta.\\
 
\subsection{Undesired effects if one does not use our constraint}

\nd If we do NOT use the constraint (\ref{ep4.8}), we have
\begin{equation}  
\label{ep4.16}
\boldsymbol{\cal P}_0=\frac {1} {2}\int |\vec{k}|\left[  
a^{+\mu v}(\vec{k})a_{\mu v}(\vec{k})-\frac {1} {2}
a_\mu^{+\mu}(\vec{k})a_v^v(\vec{k})\right]d^3k,
\end{equation}
and, appealing to the  Schwinger-Feynman variational principle we find
\[|\vec{k}|a_{\rho\lambda}^+(\vec{k^{'}})=\]
\begin{equation}
\label{ep4.17}
\frac {1} {2}\int |\vec{k}|\left\{a^{+\mu v}(\vec{k})
[a_{\mu v}(\vec{k}),a_{\rho\lambda}^+(\vec{k^{'}})]-\frac {1} {2}
a_\mu^{+\mu}(\vec{k})
[a_v^v(\vec{k}),a_{\rho\lambda}^+(\vec{k^{'}})]\right\}\;d^3k,
\end{equation}
whose solution is 
\begin{equation}
\label{ep4.18}
[a_{\mu v}(\vec{k}),a_{\rho\lambda}^{+}(\vec{k^{'}})]=
\left[\eta_{\mu\rho}\eta_{v\lambda}+\eta_{v\rho}\eta_{\mu\lambda}
-\eta_{\mu v}\eta_{\rho\lambda}\right]
\delta(\vec{k}-\vec{k^{'}}).
\end{equation}
The above is the customary graviton's quantification, that leads to a theory whose  $S$ matrix in not unitary \cite{g1,fe}.

\section{The self energy of the graviton}

\setcounter{equation}{0}

\nd To evaluate the graviton's self-energy (SF) we start with the interaction Hamiltonian   ${\cal H}_I$.
Note that the Lagrangian contains derivative interaction terms.\\
\begin{equation}
\label{ep5.1}
{\cal H}_I=\frac {\partial {\cal L}_I} {\partial\partial^0\phi^{\mu\nu}}\partial^0\phi^{\mu\nu}-{\cal L}_I.
\end{equation}
\nd A typical term reads
\begin{equation}
\label{ep5.2}
\Sigma_{G\alpha_1\alpha_2\alpha_3\alpha_4}(k)=
k_{\alpha_1} k_{\alpha_2}(\rho-i0)^{-1}\ast k_{\alpha_3}k_{\alpha_4}(\rho-i0)^{-1}.
\end{equation}
\nd where $\rho=k_1^2+k_2^2+k_3^2-k_0^2$

\nd In  $\nu$ dimensions, the Fourier transform of  (\ref{ep5.2}) is 
\[{\cal F}\{[k_{\alpha_1} k_{\alpha_2}(\rho-i0)^{-1}\ast k_{\alpha_3}k_{\alpha_4}(\rho-i0)^{-1}]_\nu\}=\]
\[-\frac {2^{2\nu-2}} {(2\pi)^\nu}\pi^\nu\left[\Gamma\left(\frac {\nu} {2}\right)\right]^2
\eta_{{\alpha_1}{\alpha_2}}\eta_{{\alpha_3}{\alpha_4}}(x+i0)^{-\nu}+\]
\[\frac {2^{2\nu-1}} {(2\pi)^\nu}\pi^\nu\Gamma\left(\frac {\nu} {2}\right)\Gamma\left(\frac {\nu} {2}+1\right)
(\eta_{\alpha_1\alpha_2}x_{\alpha_3}x_{\alpha_4}+\eta_{\alpha_3\alpha_4}x_{\alpha_1}x_{\alpha_2})
(x+i0)^{-\nu-1}-\]
\begin{equation}
\label{ep5.3}
\left.\frac {2^{2\nu}} {(2\pi)^\nu}\pi^\nu\left[\Gamma\left(\frac {\nu} {2}+1\right)\right]^2
x_{\alpha_1}x_{\alpha_2}x_{\alpha_3}x_{\alpha_4}(x+i0)^{-\nu-2}\right\}
\end{equation}
\nd where $x=x_1^2+x_2^2+x_3^2-x_0^2$

\nd Anti-transforming the above equation we have
\[[k_{\alpha_1} k_{\alpha_2}(\rho-i0)^{-1}\ast k_{\alpha_3}k_{\alpha_4}(\rho-i0)^{-1}]_\nu=\]
\[i\frac {\pi^{\frac {\nu} {2}}} {4}\frac {\left[\Gamma\left(\frac {\nu} {2}+1\right)\right]^2}
{\Gamma(\nu+2)}(\eta_{\alpha_1\alpha_2}\eta_{\alpha_3\alpha_4}+
\eta_{\alpha_2\alpha_3}\eta_{\alpha_1\alpha_4}+\eta_{\alpha_2\alpha_4}\eta_{\alpha_1\alpha_3})
\Gamma\left(-\frac {\nu} {2}\right)(\rho-i0)^{\frac {\nu} {2}}+\]
\[\left\{i\frac {\pi^{\frac {\nu} {2}}} {2}\frac {\Gamma\left(\frac {\nu} {2}\right)
\Gamma\left(\frac {\nu} {2}+1\right)}
{\Gamma(\nu+1)}(\eta_{\alpha_1\alpha_2}k_{\alpha_3}k_{\alpha_4}+
\eta_{\alpha_3\alpha_4}k_{\alpha_1}k_{\alpha_2})-\right.\]
\[i\frac {\pi^{\frac {\nu} {2}}} {2}\frac {\left[\Gamma\left(\frac {\nu} {2}+1\right)\right]^2}
{\Gamma(\nu+2)}(\eta_{\alpha_1\alpha_2}k_{\alpha_3}k_{\alpha_4}+
\eta_{\alpha_1\alpha_3}k_{\alpha_2}k_{\alpha_4}+\eta_{\alpha_1\alpha_4}k_{\alpha_2}k_{\alpha_3}+
\eta_{\alpha_3\alpha_4}k_{\alpha_1}k_{\alpha_2}+\]
\[\left.\eta_{\alpha_2\alpha_3}k_{\alpha_1}k_{\alpha_4}+\eta_{\alpha_2\alpha_4}k_{\alpha_1}k_{\alpha_3})\right\}
\Gamma\left(1-\frac {\nu} {2}\right)(\rho-i0)^{\frac {\nu} {2}-1}+\]
\begin{equation}
\label{ep5.4}
i\pi^{\frac {\nu} {2}}\frac {\left[\Gamma\left(\frac {\nu} {2}+1\right)\right]^2}
{\Gamma(\nu+2)}k_{\alpha_1}k_{\alpha_2}k_{\alpha_3}k_{\alpha_4}
\Gamma\left(2-\frac {\nu} {2}\right)(\rho-i0)^{\frac {\nu} {2}-2}
\end{equation}

\subsection{Self-Energy evaluation  for $\nu=4$}

We appeal now to a $\nu$-Laurent expansion and retain there the $\nu-4$ independent term \cite{pr}.
Thus, we Laurent-expand (\ref{ep5.4}) around  $\nu=4$ and find 

\[[k_{\alpha_1} k_{\alpha_2}(\rho-i0)^{-1}\ast k_{\alpha_3}k_{\alpha_4}(\rho-i0)^{-1}]_\nu=\]
\[i\frac {\pi^2} {\nu-4}\left\{
\frac {1} {5!}(\eta_{\alpha_1\alpha_2}\eta_{\alpha_3\alpha_4}+
\eta_{\alpha_2\alpha_3}\eta_{\alpha_1\alpha_4}+\eta_{\alpha_2\alpha_4}\eta_{\alpha_1\alpha_3})
\rho^2\right.-\]
\[\left[\frac {2} {4!}(\eta_{\alpha_1\alpha_2}k_{\alpha_3}k_{\alpha_4}+\eta_{\alpha_3\alpha_4}k_{\alpha_1}k_{\alpha_2})\right.-\]
\[\frac {1} {6!}(\eta_{\alpha_1\alpha_2}k_{\alpha_3}k_{\alpha_4}+
\eta_{\alpha_3\alpha_4}k_{\alpha_1}k_{\alpha_2}+
\eta_{\alpha_1\alpha_3}k_{\alpha_2}k_{\alpha_4}+\eta_{\alpha_1\alpha_4}k_{\alpha_2}k_{\alpha_3}+\]
\[\eta_{\alpha_2\alpha_3}k_{\alpha_1}k_{\alpha_4}+\eta_{\alpha_2\alpha_4}k_{\alpha_1}k_{\alpha_3})]
\left.\rho+\frac {8} {5!}k_{\alpha_1}k_{\alpha_2}k_{\alpha_3}k_{\alpha_4}\right\}-\]
\[\frac {i\pi^2} {5!2}
(\eta_{\alpha_1\alpha_2}\eta_{\alpha_3\alpha_4}+
\eta_{\alpha_2\alpha_3}\eta_{\alpha_1\alpha_4}+\eta_{\alpha_2\alpha_4}\eta_{\alpha_1\alpha_3})
\left[\ln(\rho-i0)+\ln\pi+C-\frac {46} {15}\right]\rho^2+\]
\[i\frac {\pi^2} {4!}\left\{(\eta_{\alpha_1\alpha_2}k_{\alpha_3}k_{\alpha_4}+
\eta_{\alpha_3\alpha_4}k_{\alpha_1}k_{\alpha_2})
\left[\ln(\rho-i0)+\ln\pi+C-\frac {8} {3}\right]\right.-\]
\[\frac {1} {24}(\eta_{\alpha_1\alpha_2}k_{\alpha_3}k_{\alpha_4}+
\eta_{\alpha_3\alpha_4}k_{\alpha_1}k_{\alpha_2}+
\eta_{\alpha_1\alpha_3}k_{\alpha_2}k_{\alpha_4}+\eta_{\alpha_1\alpha_4}k_{\alpha_2}k_{\alpha_3}+
\eta_{\alpha_2\alpha_3}k_{\alpha_1}k_{\alpha_4}+\]
\[\eta_{\alpha_2\alpha_4}k_{\alpha_1}k_{\alpha_3})
\left.\left[\ln(\rho-i0)+\ln\pi+2C-\frac {101} {15}\right]\right\}\rho-\]
\begin{equation}
\label{ep5.5}
\left.i\frac {\pi^2} {30}k_{\alpha_1}k_{\alpha_2}k_{\alpha_3}k_{\alpha_4}
\left[\ln(\rho-i0)+\ln\pi+C-\frac {47} {30}\right]+
\sum\limits_{n=1}^{\infty}a_n(\nu-4)^n\right\}.
\end{equation}
The exact value of the convolution we are interested in, i.e., the left  hand side of (5.5), is given by the  independent 
term in the above expansion, as it is well-known. If the reader is not familiar with this situation, see for instance 
 \cite{pr}. We reach
\[\Sigma_{G\alpha_1\alpha_2\alpha_3\alpha_4}(k)=
k_{\alpha_1} k_{\alpha_2}(\rho-i0)^{-1}\ast k_{\alpha_3}k_{\alpha_4}(\rho-i0)^{-1}=-\]
\[\frac {i\pi^2} {5!2}
(\eta_{\alpha_1\alpha_2}\eta_{\alpha_3\alpha_4}+
\eta_{\alpha_2\alpha_3}\eta_{\alpha_1\alpha_4}+\eta_{\alpha_2\alpha_4}\eta_{\alpha_1\alpha_3})
\left[\ln(\rho-i0)+\ln\pi+C-\frac {46} {15}\right]\rho^2-\]
\[i\frac {\pi^2} {4!}\left\{(\eta_{\alpha_1\alpha_2}k_{\alpha_3}k_{\alpha_4}+
\eta_{\alpha_3\alpha_4}k_{\alpha_1}k_{\alpha_2})
\left[\ln(\rho-i0)+\ln\pi+C-\frac {8} {3}\right]\right.-\]
\[\frac {1} {24}(\eta_{\alpha_1\alpha_2}k_{\alpha_3}k_{\alpha_4}+
\eta_{\alpha_3\alpha_4}k_{\alpha_1}k_{\alpha_2}+
\eta_{\alpha_1\alpha_3}k_{\alpha_2}k_{\alpha_4}+\eta_{\alpha_1\alpha_4}k_{\alpha_2}k_{\alpha_3}+
\eta_{\alpha_2\alpha_3}k_{\alpha_1}k_{\alpha_4}+\]
\[\eta_{\alpha_2\alpha_4}k_{\alpha_1}k_{\alpha_3})
\left.\left[\ln(\rho-i0)+\ln\pi+2C-\frac {101} {15}\right]\right\}\rho-\]
\begin{equation}
\label{ep5.6}
\left.i\frac {\pi^2} {30}k_{\alpha_1}k_{\alpha_2}k_{\alpha_3}k_{\alpha_4}
\left[\ln(\rho-i0)+\ln\pi+C-\frac {47} {30}\right]\right\}.
\end{equation}
We have to deal with  1296 diagrams  of this kind. 

\section{Including  Axions into the picture}

\setcounter{equation}{0}
Axions are  hypothetical elementary particles postulated by the Peccei–Quinn theory in 1977 to tackle  the strong CP problem in quantum chromodynamics. If they  exist and have low enough  mass (within a certain range), they could be  of interest as  possible components of cold dark matter \cite{peccei}.\vskip 3mm \nd 
  We include now a massive scalar field (axions) interacting with the graviton.
  The Lagrangian  becomes 
\begin{equation}
\label{ep6.1}
{\cal L}_{GM}=\frac {1} {\kappa^2}\boldsymbol{R}\sqrt{|g|}-\frac {1} {2}
\eta_{\mu v}\partial_\alpha h^{\mu\alpha}
\partial_\beta h^{v\beta}-\frac {1} {2} [h^{\mu v}
\partial_\mu\phi\partial_v\phi+m^2\phi^2].  
\end{equation}
We can now recast the Lagrangian in the fashion

\begin{equation}
\label{ep6.2}
{\cal L}_{GM}={\cal L}_L+{\cal L}_I+{\cal L}_{LM}+{\cal L}_{IM},
\end{equation}
where
\begin{equation}
\label{ep6.3}
{\cal L}_{LM}=-\frac {1} {2}[\partial_\mu\phi\partial^\mu\phi+m^2\phi^2],
\end{equation}
so that  ${\cal L}_{IM}$ becomes the Lagrangian for the axion-graviton action 
\begin{equation}
\label{ep6.4}
{\cal L}_{IM}=-\frac {1} {2}\kappa\phi^{\mu\nu}\partial_\mu\phi\partial_\nu\phi.
\end{equation}
 The new term in the interaction Hamiltonian is 
\begin{equation}
\label{ep6.5}
{\cal H}_{IM}=\frac {\partial {\cal L}_{IM}} {\partial\partial^0\phi}\partial^0\phi-{\cal L}_{IM}.
\end{equation}

\section{The complete Self Energy of the Graviton}

\setcounter{equation}{0}

The presence of axions generates a new contribution to the graviton's self energy
\begin{equation}
\label{ep7.1}
\Sigma_{GM\mu r v s}(k)=k_\mu k_r(\rho+m^2-i0)^{-1}\ast k_v k_s(\rho+m^2-i0)^{-1}.
\end{equation}
So as to compute it we appeal to the usual  $\nu$ dimensional integral 
together with the Feynman-parameters denoted by the letter  x.
 After a Wick rotation we obtain
\begin{equation}
\label{ep7.2}
[k_\mu k_r(\rho+m^2-i0)^{-1}\ast k_v k_s(\rho+m^2-i0)^{-1}]_\nu=
i\int\limits_0^1\int\frac {k_\mu k_r (p_v-k_v) (p_s-k_s)}
{[(k-px)^2+a]^2}d^{\nu}k dx,
\end{equation}
where

\begin{equation}
\label{ep7.3}
a=p^2x-p^2x^2+m^2.
\end{equation}
 After the variables-change $u=k-px$ we find
\begin{equation}
\label{ep7.4}
[k_\mu k_r(\rho+m^2-i0)^{-1}\ast k_v k_s(\rho+m^2-i0)^{-1}]_\nu=i
\int\limits_0^1\int\frac {f(u,x,\mu,r,v,s)}
{(u^2+a)^2}d^{\nu}u dx,
\end{equation}
where 
\[{f(u,x,\mu,r,v,s)}=u_\mu u_r p_v p_s(1-x)^2+u_\mu u_r u_v u_s-
u_\mu u_s p_r p_v x(1-x)-\]
\[u_\mu u_v p_r p_sx(1-x)-u_r u_s p_\mu p_v x(1-x)-u_r u_v p_\mu p_s x(1-x)+\]
\begin{equation}
\label{ep7.5}
p_\mu p_r p_v p_s x^2(1-x)^2+ u_v u_s p_\mu p_r x^2.
\end{equation}
 After evaluation of the pertinent integrals we arrive at 
\[[k_\mu k_r(\rho+m^2-i0)^{-1}\ast k_v k_s(\rho+m^2-i0)^{-1}]_\nu=\]
\[i\frac {(\eta_{\mu r} k_vk_s+\eta_{vs} k_\mu k_r)
m^{\nu-2}\pi^{\frac {\nu} {2}}} {8}
\Gamma\left(1-\frac {\nu} {2}\right)\times\]
\[\left[F\left(1,1-\frac {\nu} {2},\frac {3} {2};-\frac {\rho} {4m^2}\right)+\frac {1} {3}
F\left(1,1-\frac {\nu} {2},\frac {5} {2};-\frac {\rho} {4m^2}\right)\right]+\]
\[i(\eta_{\mu r}\eta_{vs}+\eta_{\mu v}\eta_{rs}+\eta_{\mu s}\eta_{vr})
\frac {\pi^{\frac {\nu} {2}}m^{\nu}} {4} 
\Gamma\left(\frac {\nu} {2}\right)
\Gamma\left(-\frac {\nu} {2}\right)
F\left(1,-\frac {\nu} {2},\frac {3} {2};-\frac {\rho} {4m^2}\right)-\]
\[i(\eta_{\mu s}k_r k_v+\eta_{\mu v}k_r k_s+\eta_{r s} k_\mu k_v+
\eta_{rv} k_\mu k_s)\frac {m^{\nu-2} \pi^{\frac {\nu} {2}}} {48}\times\]
\[\Gamma\left(1-\frac {\nu} {2}\right)
F\left(2,1-\frac {\nu} {2},\frac {5} {2};-\frac {\rho} {4m^2}\right)+\]
\begin{equation}
\label{ep7.6}
ik_\mu k_r k_v k_s\frac {m^{\nu-4}\pi^{\frac {\nu} {2}}} {12}\Gamma\left(2-\frac {\nu} {2}\right)
F\left(2,2-\frac {\nu} {2},\frac {5} {2};-\frac {\rho} {4m^2}\right).
\end{equation}

\subsection{Self-Energy evaluation  for $\nu=4$}

 We need again a Laurent's expansion and face
\[[k_\mu k_r(\rho+m^2-i0)^{-1}\ast k_v k_s(\rho+m^2-i0)^{-1}]_\nu=\]
\[-i\frac {\pi^2} {\nu-4}\left\{m^2
(\eta_{\mu r} k_vk_s+\eta_{vs} k_\mu k_r)\left[\frac {1} {3}+\frac {1} {5}
\frac {\rho} {4m^2}\right]-\right.\]
\[2{m^4}
(\eta_{\mu r}\eta_{v s}+\eta_{\mu v}\eta_{r s}+\eta_{\mu s}\eta_{rv}) \times\]
\[\left[\frac {1} {8}+\frac {1} {6}
\frac {\rho} {4m^2}+\frac {1} {15}\left(\frac {\rho} {4m^2}\right)^2\right]-\]
\[\frac {m^2} {4m^2+k^2-i0}
(\eta_{\mu s}k_r k_v+\eta_{\mu v}k_r k_s+\eta_{r s} k_\mu k_v+
\eta_{rv} k_\mu k_s)\times\]
\[\frac {k^2-m^2} {12} + \frac {m^2} {4} + \frac {k^2-m^2} {30}
\frac {\rho} {4m^2}-
\left.\frac {1} {6}k_\mu k_r k_v k_s\right\}+\]
\[i\frac {m^2\pi^2} {2}(\eta_{\mu r} k_vk_s+\eta_{vs} k_\mu k_r)\times\]
\[\left[\frac {1} {3}(\ln m^2+\ln\pi+C-1)+ 
\frac {1} {5}
\frac {\rho} {4m^2}
\left(\ln m^2+\ln\pi+C\right)\right]+\]
\[i\frac {m^2\pi^2} {30}(\eta_{\mu r} k_vk_s+\eta_{vs} k_\mu k_r)
\frac {\rho} {4m^2}\times\]
\[\left[F\left(1,1,\frac {7} {2};-\frac {\rho} {4m^2}\right)+\frac {1} {7}
F\left(1,1,\frac {9} {2};-\frac {\rho} {4m^2}\right)\right]+\]
\[-i2\pi^2m^4(\eta_{\mu r}\eta_{vs}+\eta_{\mu v}\eta_{rs}+\eta_{\mu s}\eta_{vr}) \times\]
\[\left\{\left[\frac {1} {8}-\frac {1} {6}
\frac {\rho} {4m^2}-\frac {1} {15}\left(\frac {\rho} {4m^2}\right)^2+\right]\right.\times\]
\[\left(\ln m^2+\ln\pi+1\right)-
\left.\frac {1} {2}\left[\frac {3} {32}-\frac {1} {3} \left(\frac {\rho} {4m^2}\right)\right]\right\}-\]
\[i\frac{2\pi^2m^4} {105}
(\eta_{\mu r}\eta_{vs}+\eta_{\mu v}\eta_{rs}+\eta_{\mu s}\eta_{vr})
\left(\frac {\rho} {4m^2}\right)^3
F\left(1,1,\frac {9} {2};-\frac {\rho} {4m^2}\right)-\]
\[i\frac {\pi^2m^2(k^2-m^2)} {12(4m^2+k^2-i0}
(\eta_{\mu s}k_r k_v+\eta_{\mu v}k_r k_s+\eta_{r s} k_\mu k_v+
\eta_{rv} k_\mu k_s) \times\]
\[\left[\frac {1} {2}\left(\ln m^2+\ln\pi+C-\frac {1} {4}\right)+
\frac {1} {5}\left(\ln m^2+\ln\pi+C\right)\frac {k^2} {4m^2}\right]-\]
\[i\frac {\pi^2m^2} {8(4m^2+k^2-i0}
(\eta_{\mu s}k_r k_v+\eta_{\mu v}k_r k_s+\eta_{r s} k_\mu k_v+
\eta_{rv} k_\mu k_s) \times\]
\[m^2\left[\left(\ln m^2+\ln\pi+C-\frac {1} {4}\right)+
\frac {k^2} {6}+\frac {k^2} {15}\frac {k^2} {4m^2}\right]-\]
\[i\frac {\pi^2m^2} {10}(\eta_{\mu s}k_r k_v+\eta_{\mu v}k_r k_s+\eta_{r s} k_\mu k_v+
\eta_{rv} k_\mu k_s)\times\]
\[\frac {k^2-m^2} {21(4m^2+k^2-i0)}
F\left(1,1,\frac {9} {2};-\frac {\rho} {4m^2}\right)\left(\frac {k^2} {4m^2}\right)^2-\]
\[i\frac {\pi^2} {12}k_\mu k_r k_vk_s\left[\left(\ln m^2+\ln\pi\right)+
\frac {k^2} {4m^2+k^2-i0}\right]-\]
\[i\frac {\pi^2m^2} {30}k_\mu k_r k_vk_s\frac {k^2-m^2} {4m^2+k^2-i0} 
\frac {k^2} {4m^2}F\left(1,1,\frac {7} {2};-\frac {k^2} {4m^2}\right)+\]
\begin{equation}
\label{ep7.7}
\sum\limits_{n=0}^{\infty}a_n(\nu-4)^n.
\end{equation}
 Again, the exact result for our four-dimensional convolution becomes 
\[\Sigma_{GM\mu vrs}(k)=k_\mu k_r(\rho+m^2-i0)^{-1}\ast k_v k_s(\rho+m^2-i0)^{-1}=\]
\[i\frac {m^2\pi^2} {2}(\eta_{\mu r} k_vk_s+\eta_{vs} k_\mu k_r)\times\]
\[\left[\frac {1} {3}(\ln m^2+\ln\pi+C-1)+ 
\frac {1} {5}
\frac {\rho} {4m^2}
\left(\ln m^2+\ln\pi+C\right)\right]+\]
\[i\frac {m^2\pi^2} {30}(\eta_{\mu r} k_vk_s+\eta_{vs} k_\mu k_r)
\frac {\rho} {4m^2}\times\]
\[\left[F\left(1,1,\frac {7} {2};-\frac {\rho} {4m^2}\right)+\frac {1} {7}
F\left(1,1,\frac {9} {2};-\frac {\rho} {4m^2}\right)\right]+\]
\[-i2\pi^2m^4(\eta_{\mu r}\eta_{vs}+\eta_{\mu v}\eta_{rs}+\eta_{\mu s}\eta_{vr}) \times\]
\[\left\{\left[\frac {1} {8}-\frac {1} {6}
\frac {\rho} {4m^2}-\frac {1} {15}\left(\frac {\rho} {4m^2}\right)^2+\right]\right.\times\]
\[\left(\ln m^2+\ln\pi+1\right)-
\left.\frac {1} {2}\left[\frac {3} {32}-\frac {1} {3} \left(\frac {\rho} {4m^2}\right)\right]\right\}+\]
\[i\frac{2\pi^2m^4} {105}
(\eta_{\mu r}\eta_{vs}+\eta_{\mu v}\eta_{rs}+\eta_{\mu s}\eta_{vr})
\left(\frac {\rho} {4m^2}\right)^3
F\left(1,1,\frac {9} {2};-\frac {\rho} {4m^2}\right)-\]
\[i\frac {\pi^2m^2(k^2-m^2)} {12(4m^2+k^2-i0}
(\eta_{\mu s}k_r k_v+\eta_{\mu v}k_r k_s+\eta_{r s} k_\mu k_v+
\eta_{rv} k_\mu k_s) \times\]
\[\left[\frac {1} {2}\left(\ln m^2+\ln\pi+C-\frac {1} {4}\right)+
\frac {1} {5}\left(\ln m^2+\ln\pi+C\right)\frac {k^2} {4m^2}\right]-\]
\[i\frac {\pi^2m^2} {8(4m^2+k^2-i0}
(\eta_{\mu s}k_r k_v+\eta_{\mu v}k_r k_s+\eta_{r s} k_\mu k_v+
\eta_{rv} k_\mu k_s) \times\]
\[m^2\left[\left(\ln m^2+\ln\pi+C-\frac {1} {4}\right)+
\frac {k^2} {6}+\frac {k^2} {15}\frac {k^2} {4m^2}\right]-\]
\[i\frac {\pi^2m^2} {10}(\eta_{\mu s}k_r k_v+\eta_{\mu v}k_r k_s+\eta_{r s} k_\mu k_v+
\eta_{rv} k_\mu k_s)\times\]
\[\frac {k^2-m^2} {21(4m^2+k^2-i0)}
F\left(1,1,\frac {9} {2};-\frac {\rho} {4m^2}\right)\left(\frac {k^2} {4m^2}\right)^2-\]
\[i\frac {\pi^2} {12}k_\mu k_r k_vk_s\left[\left(\ln m^2+\ln\pi\right)-
\frac {k^2} {4m^2+k^2-i0}\right]-\]
\begin{equation}
\label{ep7.8}
i\frac {\pi^2m^2} {30}k_\mu k_r k_vk_s\frac {k^2-m^2} {4m^2+k^2-i0} 
\frac {k^2} {4m^2}F\left(1,1,\frac {7} {2};-\frac {k^2} {4m^2}\right)
\end{equation}
We have to deal with 9 diagrams  of this kind.\\
\nd
Accordingly, our desired  self-energy total is a combination of 
$\Sigma_{G\alpha_1\alpha_2\alpha_3\alpha_4}(k)$ and
$\Sigma_{GM\alpha_1\alpha_2\alpha_3\alpha_4}(k)$.

\section{Self Energy of the Axion}

\setcounter{equation}{0}

Here the  self-energy  is
\begin{equation}
\label{ep8.1}
\Sigma^{\mu s} (k)=(\eta^{\mu r}\eta^{vs}+\eta^{\mu s}\eta^{vr})
k_vk_r(\rho+m^2-i0)^{-1}\ast(\rho-i0)^{-1}.
\end{equation}
In $\nu$ dimensions one has
\begin{equation}
\label{ep8.2}
[k_vk_r(\rho+m^2-i0)^{-1}\ast(\rho-i0)^{-1}]_\nu=
\int\frac {k_vk_r} {(k^2+m^2-i0)[(p-k)^2-i0]}d^\nu k.
\end{equation}
 With the  Feynman parameters used above  we obtain
\begin{equation}
\label{ep8.3}
[k_vk_r(\rho+m^2-i0)^{-1}\ast(\rho-i0)^{-1}]_\nu=
i\int\limits_0^1\int\frac {k_vk_r} {[(k-px)^2+a]^2}d^\nu kdx,
\end{equation}
 where
\begin{equation}
\label{ep8.4}
a=(p^2+m^2)x-p^2x^2.
\end{equation}
 We evaluate the integral (\ref{ep8.3}) and find

\[[k_vk_r(\rho+m^2-i0)^{-1}\ast(\rho-i0)^{-1}]_\nu=\]
\[i\frac {\eta_{vr}m^{\nu-2}\pi^{\frac {\nu} {2}}} {\nu}
\Gamma\left(1-\frac {\nu} {2}\right)
F\left(1,1-\frac {\nu} {2},\frac {\nu} {2}+1;-\frac {\rho} {m^2}\right)+\]
\begin{equation}
\label{ep8.5}
\frac {2ik_v k_rm^{\nu-4}\pi^{\frac {\nu} {2}}} {\nu+2}
\Gamma\left(2-\frac {\nu} {2}\right)
F\left(1,2-\frac {\nu} {2},\frac {\nu} {2}+2;-\frac {\rho} {m^2}\right).
\end{equation}

\subsection{Self-Energy evaluation  for $\nu=4$}

Once again, we Laurent-expand, this time (\ref{ep8.5}) around $\nu=4$, encountering 

\[[k_vk_r(\rho+m^2-i0)^{-1}\ast(\rho-i0)^{-1}]_\nu=\]
\[i\pi^2\left\{\frac {1} {\nu-4}\left(
\frac {\eta_{vr}m^2} {2}-2k_vk_r\right)\right.+\]
\[\frac {\eta_{vr}m^2} {4}\left[\left(1+\frac {1} {3}\frac {\rho} {m^2}\right)
\left(\ln m^2+\ln\pi+C-\frac {1} {2}\right)-\right.\]
\[\left.\left(1+\frac {1} {9}\frac {\rho} {m^2}\right)\right]-
\frac {k_vk_r} {3}
\left(\ln m^2+\ln\pi+C-\frac {1} {2}\right)+\]
\[\frac {1} {4}\left(\frac {\rho} {m^2}\right)\left[
\frac {\eta_{vr}m^2} {12}\frac {\rho} {m^2}-\frac {k_vk_r} {3}\right]
F\left(1,1,5;-\frac {\rho} {m^2}\right)+\]
\begin{equation}
\label{ep8.6}
\left.\sum\limits_{n=1}^{\infty}a_n(\nu-4)^n\right\}
\end{equation}
The $\nu$-independent term gives the exact convolution result we are looking for:
\[\Sigma_{vr}(k)=
k_vk_r(\rho+m^2-i0)^{-1}\ast(\rho-i0)^{-1}=\]
\[i\pi^2\left\{\frac {\eta_{vr}m^2} {4}\left[\left(1+\frac {1} {3}\frac {\rho} {m^2}\right)
\left(\ln m^2+\ln\pi+C-\frac {1} {2}\right)-\right.\right.\]
\[\left.\left(1+\frac {1} {9}\frac {\rho} {m^2}\right)\right]-
\frac {k_vk_r} {3}
\left(\ln m^2+\ln\pi+C-\frac {1} {2}\right)+\]
\begin{equation}
\label{ep8.7}
\left.\frac {1} {4}\left(\frac {\rho} {m^2}\right)\left[
\frac {\eta_{vr}m^2} {12}\frac {\rho} {m^2}-\frac {k_vk_r} {3}\right]
F\left(1,1,5;-\frac {\rho} {m^2}\right)\right\}
\end{equation}

\setcounter{equation}{0}

\section{Discussion}

\nd We have developed above the quantum field theory (QFT) of Eintein's gravity (EG), that is both unitary and finite. 
Our results critically depend on the use of a rather novel constraint the we introduced in defining  the EG-Lagrangian. 
Laurent expansions were an indispensable tool for us.
\vskip 3mm  \nd 
 In order to quantize the theory we appealed to the variational principle of Schwinger-Feynman's. This process leads to  
 just one graviton type $\phi^{12}$. \vskip 3mm  \nd
The underlying mathematics used in this effort has been developed by  Bollini et al.  \cite{tp3,tp18,tp19,tp20,pr}. This mathematics is powerful enough so as to be able to quantize 
non-renormalizable field theories  \cite{tp3,tp18,tp19,tp20,pr}.  
\vskip 3mm  \nd
We have evaluated here in finite and exact fashion, for the first time as far as we know, several quantities: 
\begin{itemize}
\item the  graviton's self-energy in the EG-field. This requires full use of the theory of distributions, 
 appealing to the possibility of creating with them a ring  with 
 divisors of zero.
\item  the above self-energy in the added presence of a massive scalar field (axions, for instance).
Two types of diagram ensue: the original ones of the pure EG field plus  the ones originated by the addition of a scalar field.
\item The axion's self-energy. 
\item Our central results revolve around Eqs. (\ref{ep5.6}) and
 (\ref{ep7.8}), corresponding to the graviton's self-energy,
 without and with the added presence of axions. Also, we give the axion's self-energy.
\end{itemize}

\end{document}